\documentclass{fmj2010}
\usepackage{graphicx}
\usepackage{hyperref}
\usepackage{amstext}
\usepackage{subfig}

\def\email#1{{e-mail: \tt#1}}

\begin{document}
   \title{Constraints on the gamma-ray emitting region in blazars from
multi-frequency VLBI measurements}

\author{K.~V.~Sokolovsky\inst{1,2}\thanks{\email{ksokolov@mpifr-bonn.mpg.de}}
          \and
           Y.~Y.~Kovalev\inst{2,1}
          \and
           A.~P.~Lobanov\inst{1}
          \and
           J.~D.~Finke\inst{3}\thanks{NRL/NRC Research Associate}
          \and
           T.~Savolainen\inst{1}
          \and
           A.~B.~Pushkarev\inst{1,4,5}
          \and
           M.~Kadler\inst{6,7,8}
          \and
           F.~K.~Schinzel\inst{1}
          \and
           V.~H.~Chavushyan\inst{9}
          \and
           L.~Carrasco\inst{9}
          \and
           A.~Carrami{\~n}ana\inst{9}
          \and
           M.~A.~Gurwell\inst{10}
          }

          \institute{Max--Planck--Institut f\"ur Radioastronomie, Auf
            dem H\"ugel 69, 53121 Bonn, Germany 
            \and
            Astro Space Center of Lebedev Physical Institute,
            Profsoyuznaya Str. 84/32, 117997 Moscow, Russia
            \and
            U.S. Naval Research Laboratory, Code 7653, 4555 Overlook 
            Ave. SW, Washington DC, 20375 USA
            \and
            Pulkovo Astronomical Observatory, Pulkovskoe Chaussee 65/1
            196140 St. Petersburg, Russia
            \and
            Crimean Astrophysical Observatory 98688 Nauchny, Crimea,
            Ukraine
            \and
            Dr. Karl Remeis-Observatory \& ECAP, Friedrich-Alexander
            University Erlangen-Nuremberg, Sternwartstr. 7, 96049
            Bamberg, Germany 
            \and 
            CRESST/NASA Goddard Space Flight Center, Greenbelt, MD
            20771, USA 
            \and 
            Universities Space Research Association, 10211 Wincopin
            Circle, Suite 500 Columbia, MD 21044, USA 
            \and 
            Instituto Nacional de Astrof\'isica, \'Optica y
            Electrnica, Apdo. Postal 51 y 216, 72000 Puebla, Pue.,
            M\'exico 
            \and 
            Harvard-Smithsonian Center for Astrophysics, Cambridge, MA
            02138, USA }

          \abstract{ Single-zone synchrotron self-Compton and external
            Compton models are widely used to explain broad-band
            Spectral Energy Distributions (SEDs) of blazars from
            infrared to gamma-rays. These models bear obvious
            similarities to the homogeneous synchrotron cloud model
            which is often applied to explain radio emission from
            individual components of parsec-scale radio jets.  The
            parsec-scale core, typically the brightest and most
            compact feature of blazar radio jet, could be the source
            of high-energy emission.  We report on ongoing work to
            test this hypothesis by deriving the physical properties
            of parsec-scale radio emitting regions of twenty bright
            Fermi blazars using dedicated $5$--$43$~GHz VLBA
            observations and comparing these parameters to results of
            SED modeling.  }

\authorrunning{Sokolovsky et al.}
\titlerunning{Constraints on the gamma-ray emitting region in blazars from multi-frequency VLBI}

   \maketitle
%


\section{Introduction}

Blazar jets are known to radiate across the entire electromagnetic
spectrum (e.g. \cite{2006AIPC..856....1M, 2009arXiv0909.2576M}).
Their radio to UV (sometimes -- X-ray) emission is believed to be
dominated by synchrotron radiation of relativistic electrons while
radiation at higher energies could be due to the inverse Compton
scattering of synchrotron photons emitted by the electrons themselves
(the synchrotron self-Compton process, SSC,
\cite{1974ApJ...188..353J,1989ApJ...340..181G}) and photons from
external sources (External Compton, EC,
\cite{1994ApJ...421..153S,2002ApJ...575..667D}).  The sources of the
external seed photons for the EC process include the accretion disc,
broad line region (BLR) clouds, warm dust and the cosmic microwave
background (CMB), with their relative contribution vary for different
blazars.

\begin{table}[t]
 \caption{Comparison of the emission region parameters estimated from the VLBA data and SED modeling}
 \centering
\[
\resizebox{\columnwidth}{!}{%
 \begin{tabular}{@{}l@{\,\,}r@{\,\,}c@{\,\,}c@{\,\,}c c@{\,\,}c@{\,\,}c@{\,\,}c@{\,\,}c@{}}
 \hline\hline
\noalign{\smallskip}
\multicolumn{5}{@{}l@{}|}{Multi-frequency VLBA }                             &                             \multicolumn{5}{@{}r@{}}{Published SED model } \\
\multicolumn{5}{@{}l@{}|}{results (this work)}                             &                             \multicolumn{5}{@{}r@{}}{parameters} \\
  Name                    & $R_{43\text{~GHz}}$   & $p^\mathrm{a,b}_\text{VLBA}$ & $B^\mathrm{b}_\text{VLBA}$    & $D_\text{var}$$^\mathrm{e}$ & $D_\text{SED}$   & $R_\text{SED}$    & $p^\mathrm{a}_\text{SED}$ & $B_\text{SED}$          & Ref.\ 
\\
                          &  {\footnotesize [$10^{15}$\,cm]} &                  &    {\small [G]}   &                             &       & {\footnotesize [$10^{15}$\,cm]} &                & {\small [G]} &           \\ 
\noalign{\smallskip}
              \hline                                                                                                
\noalign{\smallskip}
AO~0235$+$16              & $\le5800$             & $0.8$            & $\le11$           & $24.0$                      &       &                      &                &              &            \\
B0528$+$134               & $\le7600$             & $1.4$            & $\le1.2$          & $31.2$                      &       &                      &                &              & \\
S5~0716$+$714             & $\le2100$             & \multicolumn{2}{c}{$\alpha=0.4$$^\mathrm{d}$}     & $10.9$                      & $14$  & $40$                 & $2.0$          & $1$          & (1) \\
OJ~248                    & $\le6300$             & $1.6$            & $\le23$           & $13.1$                      &       &                      &                &              & \\   
OJ~287                    & $\le2800$             & \multicolumn{2}{c}{$\alpha=0.7$$^\mathrm{d}$}     & $17.0$                      &       &                      &                &              & \\
W~Com                     & $\le1700$             & $0.8$            & $\le118$          &  $1.2$                      & $20$  & $3$                  & $2.55$         & $0.35$       & (2) \\
3C~273                    & $\le1800$             & $2.0$            & $\le0.2$          & $17.0$                      & $9$   & $20$                 & $2$            & $12$         & (3) \\
3C~279                    & $\le4600$             & $1.4$            & $\le14$           & $24.0$                      &$21.5$ & $25$                 & $2.0$          & $1.8$        & (4) \\       
PKS~B1510$-$089           & $\le3900$             & \multicolumn{2}{c}{$\alpha=0.2$$^\mathrm{d}$}     & $16.7$                      & $37$  & $18$                 & $1.9$          & $0.09$       & (5) \\
4C~38.41                  & $\le3500$             & $1.0$            & $\le1.8$          & $21.5$                      &       &                      &                &              & \\
Mrk~501                   &  $\le470$             & $1.6$            & $\le67$           & $\dots$$^\mathrm{f}$        & $20$  & $1.03$               & $2$            & $0.3$        & (6) \\
NRAO~530                  & $\le6200$             & $1.6$            & $\le20$           & $10.7$                      &       &                      &                &              & \\
B1959$+$650               &  $\le510$             & $1.4$            & $\le300$          & $\dots$$^\mathrm{f}$        & $18$  & $7.3$                & $2$            & $0.25$       & (7) \\
B2155$-$304               & $\le2400$             & $1.0$            & $\le745$          & $\dots$$^\mathrm{f}$        & $32$  & $150$                & $1.3$          & $0.018$      & (8) \\
BL~Lac                    &  $\le590$             & $1.0$            & $\le3$            &  $7.3$                      &       &                      &                &              & \\
3C~454.3                  & $\le6500$             & \multicolumn{2}{c}{$\alpha=0.8$$^\mathrm{d}$}     & $33.2$                      & $26$  & $15$                 & $1.1$          & $5.4$        & (9) \\
B2344$+$514               &  $\le680$             & $1.2$            & $\le57^\mathrm{c}$& $\dots$                     &       &                      &                &              & \\
\noalign{\smallskip}
 \hline
\noalign{\smallskip}
\multicolumn{10}{@{}p{11.8cm}@{}}{%
$^\mathrm{a}$ $p$ is the power law index in the electron energy distribution $N(E)= N_0 E^{-p}$. For the optically thin part of the 
synchrotron spectrum $p = 1 - 2 \alpha$ where $\alpha$ is defined as $S_{\nu} \sim \nu^{\alpha}$. 
$^\mathrm{b}$ The estimates correspond to the parsec-scale radio core.
$^\mathrm{c}$ The value is in the observer's frame. $^\mathrm{d}$ The homogeneous synchrotron source model is not applicable for such spectrum.
$^\mathrm{e}$ The variability Doppler factor from \cite{2009A&A...494..527H} used to transform the observed magnetic field  strength 
into the source frame. 
$^\mathrm{f}$ The same Doppler factor as in the corresponding SED model was adopted.} \\
\multicolumn{10}{@{}p{11.8cm}@{}}{%
References: 
1: \cite{2008A&A...489L..37C};
2: \cite{2009arXiv0910.3750V};
3: \cite{2009A&A...494...49P};
4: \cite{2009A&A...494..509G};
5: this work;
6: \cite{2009ApJ...705.1624A};
7: \cite{2008ApJ...679.1029T}; 
8: \cite{2009ApJ...696L.150A};
9: \cite{2010arXiv1003.3476B}. } \\
\end{tabular} 
}
\]
\label{tab:one}
\end{table}

Despite the wide acceptance of the picture outlined above, the exact
location, geometry and physical properties of the regions responsible
for blazar emission at different bands remain largely unknown.
Considerable success was achieved by single-zone models (assuming a
single spherical relativistically moving blob of magnetized plasma
which emits synchrotron radiation and interacts with an ambient photon
field) in explaining blazar emission from IR to $\gamma$-ray energies
(among the many recent examples:
\cite{PKS1510FermiPaper,2010ApJ...714L.303F} and the references in
Table~\ref{tab:one}).  A significant fraction of blazar radio emission
is known to originate at the extended parsec-scale jet
and cannot be accounted for in the framework of a single-zone
model. However, emission from individual jet components resolved with
the Very Long Baseline Interferometry (VLBI) can usually be well
described by a single uniform synchrotron source model
(\cite{1970ranp.book.....P}). It is tempting to identify one of the
jet components resolved by VLBI with the zone responsible for the
emission at higher frequencies.

In this contribution we describe a program of coordinated
multi-frequency ($4.6$--$43.2$~GHz) Very Long Baseline Array (VLBA),
{\em Swift} and {\em Fermi} observations of selected $\gamma$-ray
bright blazars with the aim to identify the possible source of
high-energy radiation within the parsec-scale jet and put constraints
on physical properties based on radio spectra.

\section{Observations and analysis}

We have used the VLBA (\cite{1994IAUS..158..117N}) to observe twenty
blazars that were expected to be bright $\gamma$-ray emitters, prior
to the launch of {\em Fermi}-GST.  The observations were conducted
simultaneously at seven frequencies ($4.6$--$43.2$~GHz) utilizing the
array's unique capability of fast frequency switching between
individual VLBI scans.  After the initial calibration in {\it AIPS}
(\cite{1990apaa.conf..125G}), the sources were self-calibrated and
imaged independently at each frequency using {\it Difmap}
(\cite{Shepherd}). This software was also used to model the source
structure in the $uv$-plane and derive constraints on the core size at
$43.2$~GHz (Table~\ref{tab:one}).

A special procedure was developed to improve amplitude calibration of
the correlated flux density resulting in a $\sim 5$~\% accuracy in
$4.6$-$15.4$~GHz range and a $\sim 10$~\% accuracy at $23.8$ and
$43.2$~GHz.  Images at different frequencies were aligned with each
other using optically thin regions of the jet to compensate for phase
center shifts and possible frequency-dependent core position (``core
shift'', \cite{1998A&A...330...79L,2008A&A...483..759K}).  We applied
a spectrum extraction technique described by
\cite{2010arXiv1001.2591S} which relies on the positions and flux
densities of individual CLEAN components as opposed to the use of
restored VLBI images in order to suppress spectral artifacts arising
from the convolution with a Gaussian beam.

The sources were observed with {\em Swift} in X-ray and optical-UV
bands typically within two days after the VLBA observation. The Swift
data were analyzed using standard tools provided in the {\it
  HEAsoft}\footnote{\url{http://heasarc.nasa.gov/lheasoft/}} package.
Whenever feasible, standard stars within the UVOT field of view were
utilized to improve photometry in $UBV$ bands. The Galactic correction
was applied using the values from \cite{1998ApJ...500..525S} and
\cite{2005A&A...440..775K} for the optical-UV and X-ray absorption
respectively.

   \begin{figure}[t!]
   \centering
   \includegraphics[width=0.43\textwidth,clip=true,trim=80 30 65 80]{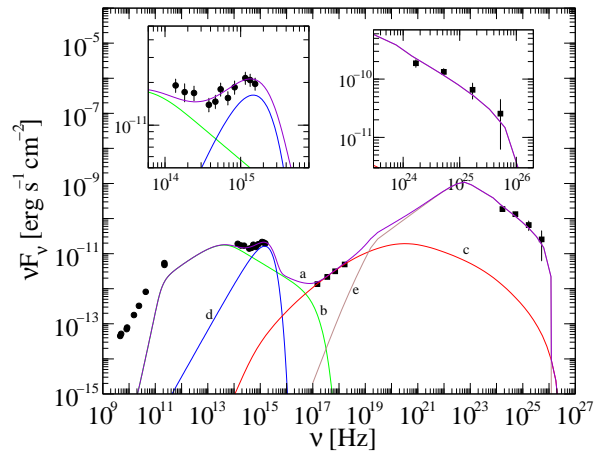}
   \caption{Quasi-simultaneous SED of PKS~B1510$-$089 constructed
     using observations with the VLBA, {\em Swift}, {\em Fermi}/LAT,
     NOT, SMA and the 2.1~m~telescope~Guillermo Haro.  The violet
     curve (a) represents the combined SED model: green curve (b) is
     the synchrotron component, red (c) is the SSC component, blue (d)
     is the accretion disk and brown (e) represents EC scattering of
     the disk radiation (see also Table~\ref{tab:1510sed}).  }
         \label{1510sed}
   \end{figure}

To obtain the quasi-simultaneous SED of PKS~B1510$-$089
(Fig.~\ref{1510sed}) discussed below, the VLBA, {\em Swift} and {\em
  Fermi}/LAT data collected during two days 2009 April 09 -- 10 were
complemented by $UBVR_ci$ photometry obtained on April 12 with the
2.56~m Nordic Optical Telescope (NOT) and infrared $JHK_s$ photometry
obtained on April 07 and 17 with the 2.1~m telescope of Guillermo Haro
Observatory in Cananea, M\'exico.  The Submillimeter Array (SMA)
provided flux density measurements at 1~mm from 2009 April 05 and
14. These observations resulted in the most well-sampled
quasi-simultaneous SED (containing multi-frequency VLBI data) among all
our sources, and for that reason it was the first we chose to
construct a model of.

$UBV$ magnitudes of PKS~B1510$-$089 observed by the {\em Swift}/UVOT
are systematically (0\hbox{$.\!\!^{\rm m}$}5--0\hbox{$.\!\!^{\rm
    m}$}3) brighter then those observed by the NOT two days
later. Since both observations were calibrated against the same set of
comparison stars from \cite{2001AJ....122.2055G}, the difference can
be interpreted as a clear sign of optical inter-day variability.

\section{Discussion of the first results}

Parsec-scale radio emission of all the observed sources
(Table~\ref{tab:one}) is dominated by a bright unresolved core (the
apparent origin of the jet, see a discussion by
\cite{2008ASPC..386..437M}) which exhibits high-amplitude flux density
variability (e.g. \cite{2010ApJ...712..405V}).  The parsec-scale core
is a natural candidate to be directly related to the bright variable
emission at $\gamma$-rays (\cite{2009ApJ...696L..17K}) and other
bands.

\begin{figure}[t!]
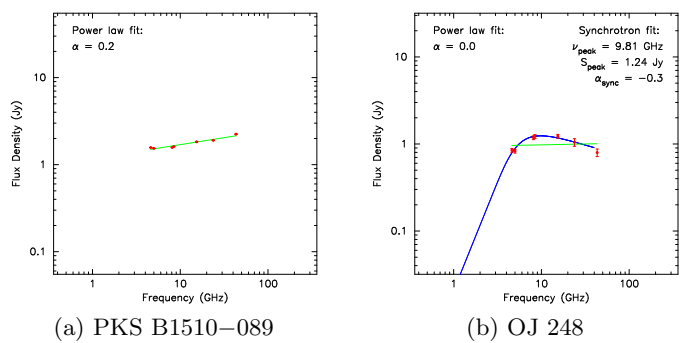

  \centering
  \subfloat[PKS~B1510$-$089]{\label{fig:1510-089spec}\includegraphics[clip,width=0.46\columnwidth]{KSokolovsky_1_Fig1a_1510-089_spectrum.eps}} \hfill
  \subfloat[OJ~248]{\label{fig:oj247spec}\includegraphics[clip,width=0.46\columnwidth]{KSokolovsky_1_Fig1b_OJ248_core_spectrum_v2.eps}}                
  \caption{
VLBA spectra of core regions of PKS~B1510$-$089 (a) and OJ~248 (b). The
  former spectrum is flat while the latter one shows a prominent synchrotron
  self-absorption peak. The green line is a power law fit. The
  blue curve is a homogeneous synchrotron source model.}
  \label{fig:sources}
\end{figure}

The multi-frequency VLBA observations have detected the synchrotron
self-absorption peak in core regions of most of the observed sources
with the exception of B0716$+$714, OJ~287, PKS~B1510$-$089 and
3C~454.3. The latter sources show inverted spectra in the
$4.6$--$43.2$~GHz range, characteristic of a partially optically-thick
inhomogeneous source.  These sources may have a detectable
self-absorption turnover at frequencies $>43$~GHz.
Fig.~\ref{fig:sources} compares core radio spectra of PKS~B1510$-$089
and OJ~248.

For sources where the self-absorption peak was detected, it was possible 
to put upper limits on the magnetic field strength
using the procedure described by \cite{2010arXiv1001.2591S}.
The source intrinsic values were computed using
variability Doppler factor estimations from \cite{2009A&A...494..527H}.
The resulting liberal upper limits on the magnetic field strength owe to
poor resolution at lower observing frequency ($4.6$~GHz) which ultimately limits
our ability to unambiguously extract spectrum of a given spatial region.

In the case of a bright isolated component, which dominates the
emission across the whole observed frequency range, it is possible to
estimate its size using the highest observing frequency ($43.2$~GHz)
instead of the lowest one and, therefore, obtain much tighter
constrains on the component size and the magnetic field strength. The
applicability of this ``isolated component'' scenario to each
individual source is currently being investigated.

Table~\ref{tab:one} presents constrains on size ($R_{43\text{~GHz}}$),
magnetic field strength ($B_\text{VLBA}$) and electron energy spectral
slope ($p_\text{VLBA}$) in the radio core region, see
\cite{2010arXiv1001.2591S} for a more detailed discussion.  These
values are compared to the corresponding parameters of SED models
($R_\text{SED}$, $B_\text{SED}$, $p_\text{SED}$) reported in the
literature.  Doppler factors from SED models ($D_\text{SED}$) and
those assumed for $B_\text{VLBA}$ computation ($D_\text{var}$) are
also listed.  
The SED model parameters are well within the constraints set by the
multi-frequency VLBA data, except for 3C\,273 which shows a
discrepancy in $B_\text{VLBA}$--$B_\text{SED}$ that may be due to
insufficient resolution of our data set. There is evidence from 86~GHz
VLBI data that $B\sim1$\,G in the mm-wavelength core of 3C\,273
(\cite{ 2008ASPC..386..451S}).

To further test the possible relation between the radio core and the
emitting blob implied by single-zone SED models, we have constructed
the quasi-simultaneous SED of PKS~B1510$-$089 presented in
Fig.~\ref{1510sed}. The use of the VLBI core flux density (instead of
the total flux density obtained by single-dish radio observations)
allows us to exclude the emission coming from the parsec-scale jet and
larger scales. Our ambition was to check how this will affect the SED
modeling.

\begin{table}[h]
 \centering
 \caption{PKS~B1510$-$089 SED model parameters}
 \begin{tabular}{@{}l r@{\,=\,} l@{}}
 \hline\hline
\noalign{\smallskip}
 Minimum $e^-$ Lorentz factor & $\gamma_\text{min}$ & $30$    \\
 Brake Lorentz factor & $\gamma_\text{brk}$  & $1.9 \times 10^3$ \\
 Maximum $e^-$ Lorentz factor & $\gamma_\text{max}$ & $1.0 \times 10^5$ \\
 $e^-$ energy slope below $\gamma_\text{brk}$ & $p_1$     & $1.9$   \\
 $e^-$ energy slope above $\gamma_\text{brk}$ & $p_2$     & $3.9$   \\
 Doppler factor & $D$                 & $37$    \\
 Bulk Lorentz factor & $\Gamma$            & $37$    \\
 Magnetic field strength & $B$                 & $0.09$~G \\
 Variability time  & $t_\text{var}$      & $2.16 \times 10^4$~sec \\
 Blob radius & $R$                 & $1.8 \times 10^{16}$~cm \\
 Jet power (magnetic field) & $P_\text{jet, B} $ & $2.8 \times 10^{43}$~erg/s \\
 Jet power (electrons)      & $P_{\text{jet}~e^-} $ & $7.5 \times 10^{45}$~erg/s \\
 Black hole mass  & $M_\text{BH} $  & $1 \times 10^9 M_\odot$ \\
 Accretion efficiency & $\eta  $ & 1/12 \\
 Eddington ratio & ${ L_\text{disk} }/{ L_\text{Edd} }$ & $0.06$ \\
 Blob distance from disk & $r $ & $5.1 \times 10^{17}$~cm \\
\noalign{\smallskip}
 \hline
 \end{tabular} 
 \label{tab:1510sed}
\end{table}

Contribution from a hot thermal component, probably the accretion
disk, is evident in the optical and UV bands. The SED in the IR to
$\gamma$-ray range can be explained by the single zone EC model with
the Compton scattered accretion disk photons being responsible for the
bulk of the observed $\gamma$-ray emission. In this model, the blob is
close enough to the accretion disk so the accretion disk photons can
reach the blob directly, with no need to be scattered back from BLR
clouds or dust. Note, that due to relativistic aberration some of the
disk photons will be coming to the blob (nearly) head-on.  The photons
will primarily come from a radius of the disk of $\sqrt{3} r$ where r
is the blob's distance from the black hole
(\cite{2002ApJ...575..667D}).  The blob size was constrained by the
observed $\gamma$-ray variability time scale of the order of a few
hours (\cite{PKS1510FermiPaper}).  Parameters of the model are
summarized in Table~\ref{tab:1510sed}.
We note that the model fit is not unique, single-zone models with a
different source of seed photons could probably also produce an
acceptable fit (cf. \cite{PKS1510FermiPaper}).

The constructed model (Fig.~\ref{1510sed}) represents well the IR to
$\gamma$-ray data, however it dramatically underpredicts the radio
emission because of the synchrotron self-absorption occurring in the
small blob at a high frequency. A significantly larger blob size would
be inconsistent with the observed variability timescale.  Even if the
variability timescale argument were dismissed, a single-zone SED model
with a larger blob would have difficulties in explaining the observed
hard radio spectrum which is inconsistent with the softer X-ray
spectrum or optically thick synchrotron radiation from a uniform
source.  One way to overcome this difficulty would be to introduce a
second brake in the electron energy spectrum.

Alternatively, one could abandon the attempt to explain the whole SED
with a single-zone model and assume that even the radio emission
observed from the cm-mm band core is coming from a larger
($R_\text{SED} < R < R_{43\text{~GHz}}$) structure downstream of the
blob responsible for the IR-to-$\gamma$-ray emission. This structure
(which cannot be observed separately from the blob due to limited
resolution of the available VLBI data) could be a smooth jet
(\cite{1979ApJ...232...34B}) or a number of distinct jet components
(\cite{1980Natur.288...12M}) -- perhaps the blobs which contributed to
high-energy emission earlier... One could argue that currently this
extended region does not contribute significantly to the optical and
$\gamma$-ray emission on the basis of the observed short timescale
variability in these bands. A more detailed SED modeling is needed to
test this scenario.

\section{Summary}

The program of coordinated multi-frequency ($4.6$--$43.2$~GHz)
VLBA, {\em Swift} and {\em Fermi} observations of
selected $\gamma$-ray bright blazars is described and the first analysis
results are discussed.
Constraints on the size ($R_{43\text{~GHz}}$) and magnetic field strength
($B_\text{VLBA}$) in the parsec-scale core region derived from the VLBA
observations are consistent with the values ($R_\text{SED}$, $B_\text{SED}$)
usually assumed in single-zone SED models (Table~\ref{tab:one}).
However, the single-zone SED model which
we tested for PKS~B1510$-$089 has difficulties in explaining the observed level
of core radio emission while being consistent with the observed $\gamma$-ray and
optical variability timescales.
The multi-frequency VLBA and SED data analysis for the whole sample is in progress.
The SED data on PKS~B1510$-$089 discussed here are available from
KVS upon request.

\begin{acknowledgements}
  KVS is supported by the International Max-Planck Research School
  (IMPRS) for Astronomy and Astrophysics at the universities of Bonn
  and Cologne. YYK was supported in part by the return fellowship of
  Alexander von Humboldt foundation and the Russian Foundation for
  Basic Research (RFBR) grant 08-02-00545. This work is based on data
  obtained from the National Radio Astronomy Observatory's Very Long
  Baseline Array (VLBA), project BK150. The National Radio Astronomy
  Observatory is a facility of the National Science Foundation
  operated under cooperative agreement by Associated Universities,
  Inc. The Submillimeter Array is a joint project between the
  Smithsonian Astrophysical Observatory and the Academia Sinica
  Institute of Astronomy and Astrophysics and is funded by the
  Smithsonian Institution and the Academia Sinica. The data presented
  here have been taken using ALFOSC, which is owned by the Instituto
  de Astrofisica de Andalucia (IAA) and operated at the Nordic Optical
  Telescope under agreement between IAA and the NBIfAFG of the
  Astronomical Observatory of Copenhagen. The authors acknowledge the
  support by the staff of the Observatorio Astrof\'isico Guillermo Haro.
\end{acknowledgements}

\end{document}